\documentclass{article}
\usepackage{spconf}

\usepackage[T1]{fontenc}
\usepackage{cite}
\usepackage{amsmath,amssymb,amsfonts}
\usepackage{xcolor}
\usepackage{amsmath, bm}
\usepackage{amssymb}

\usepackage{algorithm}
\usepackage{algpseudocode}

\usepackage{graphicx}
\usepackage[nolist,printonlyused]{acronym}
\usepackage{times}
\usepackage{subcaption}

\DeclareMathOperator{\EX}{\mathbb{E}}

\newcommand\s{\bm{s}}
\newcommand\n{\bm{n}}
\newcommand\y{\bm{y}}
\newcommand\x{\bm{x}}

\newcommand\g{\bm{g}}

\newcommand\A{\bm{A}}
\newcommand\B{\bm{B}}

\newcommand\K{\bm{K}}
\renewcommand\H{\bm{H}}
\newcommand\F{\bm{F}}
\newcommand\W{\bm{W}}
\newcommand\G{\bm{G}}
\newcommand\I{\bm{I}}
\newcommand\U{\bm{U}}

\newcommand\Bt{\bm{\mathcal{B}}}
\newcommand\Gt{\bm{\mathcal{G}}}

\newcommand\UE{\mathrm{MS}}
\newcommand\FD{\mathrm{FD}}
\newcommand\Fro{\mathrm{F}}
\newcommand\RS{\mathrm{RS}}

\newcommand\Rx{\mathrm{R}}
\newcommand\Tx{\mathrm{T}}

\newcommand\diag{\mathrm{diag}}

\newcommand\CC{\mathbb{C}}

\title{Title of the paper}
\name{Sepideh Gherekhloo, Khaled Ardah, and Martin Haardt
	\thanks{The authors gratefully acknowledge the support of the German Research Foundation (DFG) under contracts no. HA 2239/14-1 and no. HA 2239/6-2.}
}
\address{Communications Research Laboratory (CRL), TU Ilmenau, Ilmenau, Germany \\
	\{sepideh.gherekhloo, khaled.ardah, martin.haardt\}@tu-ilmenau.de }

\begin{document}
	
	\title{{ { Fully Digital and  Hybrid Beamforming Design For Millimeter-Wave MIMO-OFDM Two-Way Relaying Systems}}\\}
	
	
	\maketitle
	
	\begin{abstract}
		In this work, we consider the design of hybrid analog-digital (HAD) multi-carrier MIMO-OFDM two-way relaying systems, where the relay station is equipped with a HAD amplify-and-forward architecture and every mobile station is equipped with a fully-digital beamforming architecture. We propose a sub-optimal solution by reformulating the original non-convex problem as a constrained Tucker2 decomposition with the objective of minimizing the sum Euclidean-norm between the HAD amplification matrices and their fully-digital counterparts. For the fully-digital amplification matrix design, we use a Frobenius-norm maximization of the effective channels on every subcarrier and propose an effective solution applicable for multi-stream communication scenarios. After that, we propose an alternating maximization (AltMax) HAD solution by exploiting the tensor structure of the reformulated problem. Simulation results are provided, where we show that the proposed fully-digital and AltMax-based HAD amplification matrix designs outperform some benchmark methods, especially for multi-stream communication scenarios.

	\end{abstract}
	
	\begin{keywords}
	Two-way relaying, hybrid beamforming, millimeter-wave communication, Tucker2 tensor decomposition.
	\end{keywords}
	
	\vspace{-10pt}
	\section{Introduction}
\vspace{-5pt}
Recently, the integration of millimeter-wave (mm-wave) frequencies and massive MIMO techniques has been recognized as one of the key solutions to improve the spectral efficiency (SE) of 5G and beyond mobile systems \cite{Larsson2014,Heath_2016}. However, two major practical challenges arise. On the one hand, the conventional fully-digital beamforming architectures become impractical, due to their high-cost and high-energy consumption, as they require a dedicated radio-frequency (RF) chain for each antenna. On the other hand, high-frequency communications are very sensitive to signal-blockage and signal-attenuation, which makes mm-wave-based systems mainly applicable to short-range communications. Therefore, hybrid analog-digital (HAD) beamforming architectures \cite{Heath_2016,Rial2016,Ardah2020,Gherekhloo2020,Ardah2018} and relay-assisted communications \cite{Unger2008,Roemer2009,Roemer2010b,Sun2015,Sharma2018,Chauhan2018} have been proposed to tackle these challenges. In HAD architectures, the number of RF chains can be reduced significantly, compared to the number of the antenna elements. Meanwhile, relay stations (RSs) can be used to increase the communication range.

Among several relaying schemes, two-way relaying (TWR) is known to be more resources efficient \cite{Roemer2009}. In TWR \cite{Unger2008,Roemer2009,Roemer2010b,Roemer2010,Sun2015,Sharma2018,Chauhan2018}, the data transmission between two mobile stations (MSs) occurs in two subsequent transmission phases. First, both MSs transmit to the RS. Then, the RS transmits back to both MSs, either using a decode-and-forward (DF) or an amplify and forward (AF) relaying architecture. In this paper, we focus our attention on the AF scheme. Moreover, it is known that TWR systems can be transformed into two separate single-user MIMO channels if the self-interference at every MS is subtracted \cite{Roemer2009,Roemer2010b}. This implies that, if the amplification matrix is known, the optimal decoding and encoding matrices at the MSs are given by the dominant singular vectors of the resulting effective channels. Thus, the remaining challenge is how to design the RS amplification matrix. To this end, the authors of \cite{Roemer2009} proposed a method, called Algebraic norm-maximizing (ANOMAX), which designs the RS amplification matrix that maximizes the weighted sum of the Frobenius norms of the effective channels. However, based on the observations, ANOMAX is mainly applicable to single-stream transmissions. The authors of \cite{Roemer2010b} extended ANOMAX by proposing rank-restored ANOMAX (RR-ANOMAX) to increase the rank of the amplification matrix by adjusting its singular values while preserving its subspaces. On the other hand, the authors of \cite{Unger2008} proposed an amplification matrix design using the zero-forcing (ZF) and minimum mean square error (MMSE) filters. However, it was reported in \cite{Roemer2009}, using computer simulations, that the ANOMAX-based solutions outperform the ZF-based and the MMSE-based solutions. 

In this paper, we develop a new fully-digital RS amplification matrix design for wide-band multi-carrier MIMO OFDM AF TWR systems, which extends the results of \cite{Roemer2009} and \cite{Roemer2010b}. To this end, we consider the Frobenius-norm maximization of the effective channels  on every subcarrier and propose a different solution than \cite{Roemer2010b} 
to further enhance the rank of the obtained fully-digital RS amplification matrices, termed enhanced RR-ANOMAX (ERR-ANOMAX). As the second contribution of the paper, we design HAD amplification matrix by reformulating the original non-convex problem as a constrained Tucker2   decomposition \cite{Tucker1966,HOSVDLathauwer,Cichocki2015} with the objective of minimizing Euclidean-norm between the HAD amplification matrices and their fully-digital counterparts. Then, we propose two HAD solutions, a non-iterative HOSVD-based solution \cite{HOSVDLathauwer} and an iterative alternating maximization (AltMax)-based solution \cite{Gherekhloo2020}. The simulation results show that ERR-ANOMAX outperforms ANOMAX and RR-ANOMAX in multi-stream communication scenarios.

\vspace{-10pt}
\section{System Model}\label{sec:Section1}
\begin{figure}
	\centering
	\includegraphics[width=0.7\linewidth,trim = {0cm 0cm 0cm 0cm}]{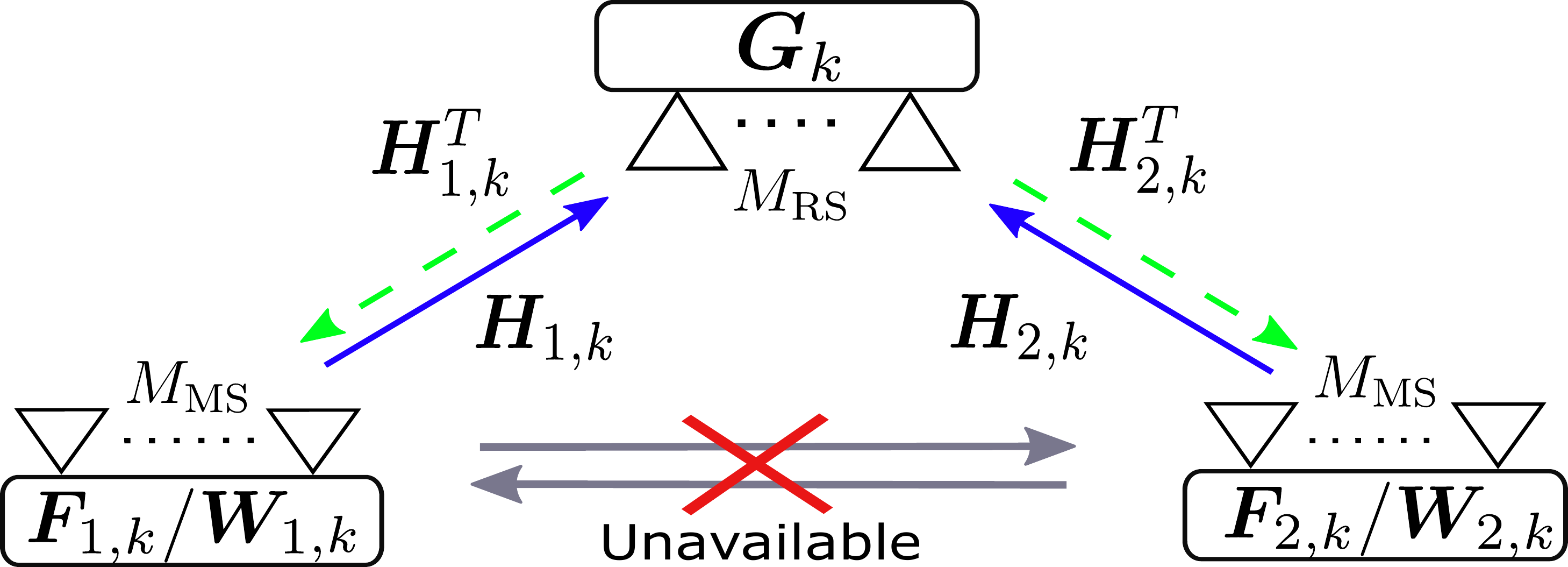}
	\caption{An AF TWR system.}
	\label{fig:TWR}
	\vspace{-10pt}
\end{figure}

\vspace{-5pt}
As depicted in Fig. \ref{fig:TWR}, we consider\footnote{Notation: The transpose, the Hermitian transpose, the Moore-Penrose pseudo-inverse, the Kronecker product, and the Khatri-Rao product are denoted by $\bm{X}^{T}, \bm{X}^{H}, \bm{X}^+, \otimes, \text{and}~ \diamond$, respectively. The Frobenius norm is denoted by $\|\cdot\|_\Fro$. The operator $\diag\{\bm{x}\}$ returns a diagonal matrix with $\bm{x}$ as the diagonal arguments, $\text{vec}\{\bm{X}\}$ stacks the columns of matrix $\bm{X}$ into a vector, and $\text{unvec}\{\bm{x}\}$ is the inverse of the $\text{vec}\{\bm{X}\}$ operator. The following property is also used:  $\text{vec}\{\bm{A}  \diag(\bm{b}) \bm{C}\} = (\bm{C}^{T} \diamond \bm{A}) \bm{b}$.} an AF TWR mm-wave massive MIMO-OFDM communication system, where two MSs communicate with each other via an intermediate RS over $K$ subcarriers.   It is assumed that the RS and MS $\ell$ are equipped with $M_{\RS}$ and $M_{\ell}$ antennas, respectively. For $\ell\in \{1,2\}$ and $ k\in \{0,\dots,K-1\}$, let $\H_{\ell,k}$ be $M_{\RS} \times M_{\ell}$ MIMO channel between MS $\ell$ and the RS on the $k$th subcarrier. We assume a TDD system so that the channel reciprocity holds, i.e., the reverse (downlink) channels are the transpose of the forward (uplink) channels. Let  $\F_{\ell,k} $ and $\W_{\ell,k}$ denote the $M_{\ell} \times N_{s}$ fully-digital precoding and decoding matrices of MS $\ell$ on the $k$th subcarrier, respectively, where $N_s$ denotes the number of data streams. Moreover, let $\G_{k}$ denote the $M_{\RS} \times M_{\RS}$ RS amplification matrix on the $k$th subcarrier.

In TWR systems, the data transmission occurs in two phases. In phase 1, the MSs transmit their signals to the RS simultaneously so that the received signal at the RS on the $k$th subcarrier can be written as
\begingroup\makeatletter\def\f@size{9.5}\check@mathfonts
\def\maketag@@@#1{\hbox{\m@th\small\normalfont#1}}%
\begin{equation}
	\x_{k} = \H_{1,k} \F_{1,k} \s_{1,k} + \H_{2,k} \F_{2,k} \s_{2,k} + \n_{\RS,k} ,
	\label{eq:RecAtRelay}
\end{equation} 
\endgroup
where $\s_{\ell,k}$ is the  data vector with $\mathbb{E}\{\s_{\ell,k} \s^{H}_{\ell,k}\}   = \bm{I}_{N_{s}}$ and $\n_{\RS,k}$ contains zero-mean circularly symmetric complex Gaussian noise with variance $\sigma^2_{\RS}$. Here, it is assumed that $\EX\{\Vert \F_{\ell,k}\Vert_{2}^2 \} = P_{\UE}$, where $P_{\UE}$ denotes the maximum transmit power of a MS per subcarrier. In phase 2, the RS transmits $\x_{k}$ to both MSs so that the received signal  at MS~$\ell$ on the $k$th subcarrier is given as 
\begingroup\makeatletter\def\f@size{9.5}\check@mathfonts
\def\maketag@@@#1{\hbox{\m@th\small\normalfont#1}}%
\begin{align}\label{RxSigk}
	\y_{\ell,k}  = & \W^{H}_{\ell,k}  {\H^{T}_{\ell,k}} \G_{k} \x_{k} + \W^{H}_{\ell,k}  \tilde{\bm{n}}_{\ell,k},
\end{align}
\endgroup
where $\tilde{\bm{n}}_{\ell,k} = {\H^{T}_{\ell,k}} \G_{k} \n_{\RS,k} + \n_{\ell,k}$ is the total additive noise, in which $\n_{\ell,k}$ denotes the additive white Gaussian noise at MS $\ell$ with variance $\sigma^2_{\UE}$. Similarly, we assume that $\EX\{\Vert \G_{k} \x_{k} \Vert_{2}^2 \} = P_{\RS}$, where $P_{\RS}$ denotes the RS maximum transmit power per subcarrier. By expanding (\ref{RxSigk}), we have
\begingroup\makeatletter\def\f@size{9.5}\check@mathfonts
\def\maketag@@@#1{\hbox{\m@th\small\normalfont#1}}%
\begin{align}\label{RxSigk3}
	\y_{\ell,k}  = \y^{\text{DS}}_{\ell,k}  + \y^{\text{SI}}_{\ell,k}  +   {\W^{H}_{\ell,k}} \tilde{\bm{n}}_{\ell,k},
\end{align}
\endgroup
where $ \y^{\text{SI}}_{\ell,k} = {\W^{H}_{\ell,k}}  \bar{\bm{H}}_{\ell,\ell,k} \F_{\ell,k} \s_{\ell,k}$ is the self-interference (SI) and $\y^{\text{DS}}_{\ell,k} = {\W^{H}_{\ell,k}} \bar{\bm{H}}_{\ell,\ell^{\prime},k} \F_{\ell^{\prime},k} \s_{\ell^{\prime},k},\ell\neq \ell^{\prime}$, is the desired-signal (DS),
in which we have defined the effective channels
\begingroup\makeatletter\def\f@size{9.5}\check@mathfonts
\def\maketag@@@#1{\hbox{\m@th\small\normalfont#1}}%
\begin{align}\label{Hef}
	\bar{\bm{H}}_{\ell,\ell^{\prime},k} =  {\H^{T}_{\ell,k}} \G_{k} \H_{\ell^{\prime},k},  ~\ell,\ell^{\prime} \in \{1,2\}.
\end{align} \endgroup

Note that the SI appears in (\ref{RxSigk3}) since each MS receives back its own signal from the RS. However, under ideal conditions, as elaborated in \cite[Section III-B]{Roemer2009}, the SI term can be subtracted completely by MSs. Consequently, (\ref{RxSigk3}) reduces to
\begingroup\makeatletter\def\f@size{9.5}\check@mathfonts
\def\maketag@@@#1{\hbox{\m@th\small\normalfont#1}}%
\begin{align}\label{RxSigk2}
	\bm{z}_{\ell,k}  = \y_{\ell,k} - \y^{\text{SI}}_{\ell,k}  = \y^{\text{DS}}_{\ell,k} + \W^{H}_{\ell,k} \tilde{\bm{n}}_{\ell,k}.
\end{align} 
\endgroup

From the above, the SE of MS~$\ell$ on the $k$th subcarrier can be expressed as 
\begingroup\makeatletter\def\f@size{9.5}\check@mathfonts
\def\maketag@@@#1{\hbox{\m@th\small\normalfont#1}}%
\begin{equation}\label{SEk}
	\textrm{SE}_{\ell,k} = \frac{1}{2} \log_2 \Big| \Big (\I_{N_s} + \Big(\W^{H}_{\ell,k} \bm{\Phi}_{\ell,k} \W_{\ell,k}\Big)^{-1} \bm{R}_{\ell,k} \Big|,
\end{equation} \endgroup
where \begingroup\makeatletter\def\f@size{9}\check@mathfonts
\def\maketag@@@#1{\hbox{\m@th\small\normalfont#1}}$\bm{\Phi}_{\ell,k} = \mathbb{E} \{\tilde{\bm{n}}_{\ell,k}  \tilde{\bm{n}}^{H}_{\ell,k}  \} =  {\sigma^2_{\RS} \H^{T}_{\ell,k}} \G_{k} \G^{H}_{k} {\H^{*}_{\ell,k}} + \sigma^2_{\UE} \bm{I}_{M_{\UE}}$, \endgroup  \begingroup\makeatletter\def\f@size{9}\check@mathfonts
\def\maketag@@@#1{\hbox{\m@th\small\normalfont#1}}$\bm{R}_{\ell,k} = \mathbb{E} \{ \big( \y^{\text{DS}}_{\ell,k}  \big) \big( \y^{\text{DS}}_{\ell,k} \big)^{H} \} = {\W^{H}_{\ell,k}} \bar{\bm{H}}_{\ell,\ell^{\prime},k} \F_{\ell^{\prime},k} \F^{H}_{\ell^{\prime},k} \bar{\bm{H}}^{H}_{\ell,\ell^{\prime},k} {\W_{\ell,k}}$ \endgroup are the effective noise and the desired signal covariance matrices, respectively. 

Our goal is to design the beamforming matrices to maximize the total SE of the system, which can be expressed as
\begin{equation}
	\begin{aligned}
		&\max_{ \{ \F_{\ell,k}, \W_{\ell,k}, \G_{k}, \forall k \} }  \quad  \textstyle \sum_{\ell= 1}^2 \sum_{k = 0}^{K - 1}\textrm{SE}_{\ell,k}\\ 
		\textrm{s.t.} & \quad \|\F_{\ell,k}\|^2_{\Fro} = P_{\UE},  
		\quad \|\G_{k} \x_{k}\|^2_{2} = P_{{\RS}}.
		\label{eq:SE}
	\end{aligned}
\end{equation}

Note that (\ref{eq:SE}) is a non-convex optimization function due to the joint optimization of the beamforming matrices at the MSs and the amplification matrix at the RS. To obtain a solution, we propose in the following a sub-optimal non-iterative approach by decoupling the optimization procedure between the beamforming matrices of the MSs and the RS amplification matrix. By noting that for any given $\G_{k}$ on subcarrier~$k$, the system model in (\ref{RxSigk2}) reduces to two independent point-to-point MIMO communication systems, one for each MS. It implies that the optimal precoding (resp. decoding) matrix, on every subcarrier, is given by the dominant $N_s$ right (resp. left) singular vectors of the respective effective channel given by (\ref{Hef}), with the power allocation given by the water-filling method \cite{Fonollosa2005}. Specifically, by dropping the subcarrier index $k$, to ease the notation, we assume that MS $\ell$ first calculates the noise whitening filter $\bm{Q}_{\ell} = \bm{\Phi}^{-{1}/{2}}_{\ell}$ so that $\bm{Q}_{\ell} \bm{\Phi}_{\ell} \bm{Q}_{\ell} = \bm{I}_{M_\UE}$. Let $\widetilde{\bm{H}}_{\ell,\ell^\prime}  = \bm{Q}_{\ell} \bar{\bm{H}}_{\ell,\ell^\prime}$ be the whitened effective channel between  MS $\ell$ and  MS $\ell^{\prime}$, with $\ell \neq \ell^{\prime}$. Then, given the SVD of $ \widetilde{\bm{H}}_{\ell,\ell^\prime} = \widetilde{\bm{U}}_{\ell,\ell^\prime} \text{diag}\{\widetilde{\bm{\lambda}}_{\ell,\ell^\prime}\}  \widetilde{\bm{V}}^{H}_{\ell,\ell^\prime}$, the precoding matrix  of MS~$\ell^{\prime}$ is given as $\F_{\ell^{\prime}} = [\widetilde{\bm{V}}_{\ell,\ell^\prime}]_{[1:N_s]}$, while the decoding matrix of the MS~$\ell$ is given as ${\W_{\ell}} = [\widetilde{\bm{U}}_{\ell,\ell^\prime}]_{[1:N_s]}$. Accordingly, (\ref{SEk}) can be simplified as
\begingroup\makeatletter\def\f@size{9.5}\check@mathfonts
\def\maketag@@@#1{\hbox{\m@th\small\normalfont#1}}%
\begin{align}\label{SEk2}
	\text{SE}_{\ell} = \textstyle\sum_{i = 1}^{N_s} \log_2 (1 + ([\widetilde{\bm{\lambda}}_{\ell,\ell^\prime}]_{[i]})^2 \cdot p_{\ell,\ell^\prime,i}),
\end{align} \endgroup
where ${p_{\ell,\ell^\prime,i}}$ is the power of the $i$th data stream given by ${p_{\ell,\ell^\prime,i}} = \max \big\{ {1}/{\mu_{\ell,\ell^\prime}} - {1}/{[\widetilde{\bm{\lambda}}_{\ell,\ell^\prime}]_{[i]}}   ,0 \big\}$, where $\mu_{\ell,\ell^\prime}$ is the real-valued water-level found using, e.g., the bisection method so that $\sum_{i = 1}^{N_s} {p_{\ell,\ell^\prime,i}} = P_{\UE}$. From the above, the problem that we address in this paper is the design of $\G_{k},\forall k$.

\vspace{-10pt}
\section{Fully-digital-based design}\label{FDRS}
\vspace{-5pt}
In this  section, we drop the subcarrier index $k$, since the proposed fully-digital design is decoupled between the $K$ subcarriers. From (\ref{SEk2}), it is clear that the RS amplification matrix $\G$, of any subcarrier, should be designed so that the singular values of the whitened effective channels of both MSs, i.e., $\widetilde{\bm{\lambda}}_{1,2}$ and $\widetilde{\bm{\lambda}}_{2,1}$, are maximized. To achieve this goal, the authors in \cite{Roemer2009} proposed a heuristic closed-form approach, termed ANOMAX, to design $\G$ as
\begingroup\makeatletter\def\f@size{9.5}\check@mathfonts
\def\maketag@@@#1{\hbox{\m@th\small\normalfont#1}}%
\begin{align}\label{p1}
	\G =  &\max_{\G} \Big( \big\Vert \bar{\bm{H}}_{1,2} \big\Vert^{2}_{F} + \big\Vert  \bar{\bm{H}}_{2,1} \big\Vert^{2}_{F} \Big).
\end{align} 
\endgroup
After some algebraic manipulations, (\ref{p1}) is rewritten as  \cite{Roemer2009} 
\begingroup\makeatletter\def\f@size{9.5}\check@mathfonts
\def\maketag@@@#1{\hbox{\m@th\small\normalfont#1}}%
\begin{align}\label{p2}
	\g =  &\max_{\g}  \Vert \bm{K} \g \Vert^{2}_2, \quad  \text{ s.t. } \quad  \Vert \g \Vert_{2} = 1,
\end{align} \endgroup
where $\bm{K} = [( \bm{H}_2 \otimes  \bm{H}_1 ) ,  ( \bm{H}_1 \otimes  \bm{H}_2 ) ]^{T} $ and $\g = \text{vec}\{\G\}$. Using the SVD  of $\bm{K} = \bm{U}_{\K} \text{diag}(\bm{\lambda}_{\K}) \bm{V}^{H}_{\K}$, the optimal solution of (\ref{p2}) is given as $\g_{\text{ANOMAX}} = [\bm{V}_{\K}]_{[:,1]}$. However, it was observed in \cite{Roemer2009} that the unfolded matrix, i.e., $\G_{\text{ANOMAX}} = \text{unvec}\{ \g_{\text{ANOMAX}} \}$ often exhibits a low-rank structure. To make it applicable for multi-stream communications, the authors extended their framework in \cite{Roemer2010b} by proposing RR-ANOMAX. In RR-ANOMAX, the rank of $\G_{\text{ANOMAX}}$ is increased by adjusting its singular values while preserving its subspaces. In the following, we build on those results and propose a method that further enhances the rank of $\G_{\text{ANOMAX}}$, termed hereafter enhanced RR-ANOMAX (ERR-ANOMAX), by exploiting the structure of problem (\ref{p2}).

In contrast to ANOMAX and RR-ANOMAX, in ERR-ANOMAX, the $\g$ vector is constructed 
by taking the contributions of the first $R$ vectors in $\bm{V}_{\K}$ as
\begingroup\makeatletter\def\f@size{9.5}\check@mathfonts
\def\maketag@@@#1{\hbox{\m@th\small\normalfont#1}}%
\begin{align}\label{ge}
	\g_{\text{E-ANOMAX}} =  [\bm{V}_{\K}]_{[:,1]} + \dots + [\bm{V}_{\K}]_{[:,R]} ,
\end{align} \endgroup
where $R$ is a design parameter that we investigate numerically in Section \ref{numresults}. Then, for the unfolded matrix $\G_{\text{E-ANOMAX}} = \text{unvec}\{ \g_{\text{E-ANOMAX}} \}$, the SVD is given as $\G_{\text{E-ANOMAX}} = \bm{U}_{\G} \text{diag}(\bm{\lambda}_{\G}) \bm{V}^{H}_{\G}$. This implies that $\g_{\text{E-ANOMAX}} =  (\bm{V}^{*}_{\G}  \diamond \bm{U}_{\G}) \bm{\lambda}_{\G}$. Here, we propose to redesign $\bm{\lambda}_{\G}$ while keeping both $\bm{U}_{\G}$ and $\bm{V}_{\G}$ fixed, to enhance the rank of $\G_{\text{E-ANOMAX}}$. Specifically, by substituting $\g_{\text{E-ANOMAX}}$ into (\ref{p2}), we have
\begin{align}\label{p3}
	\bm{\lambda}^{\star}_{\G} =  &\max_{\bm{\lambda}_{\G}  }  \Vert \tilde{\bm{K}} \bm{\lambda}_{\G} \Vert^{2}_2 \quad  \text{ s.t. } \quad  \Vert \bm{\lambda}_{\G} \Vert_{2} = 1,
\end{align} 
where $\tilde{\bm{K}} = \bm{K} (\bm{V}^{*}_{\G}  \diamond \bm{U}_{\G})$. 
In contrast to (\ref{p2}), the optimization variable of (\ref{p3}) is a real-valued vector. Therefore, we propose to design $\bm{\lambda}_{\G}$ using a water-filling like method \cite{Fonollosa2005}. Specifically, given the SVD of $\tilde{\bm{K}} = \bm{U}_{\tilde{\bm{K}}} \text{diag}(\bm{\lambda}_{\tilde{\bm{K}}}) \bm{V}^{H}_{\tilde{\bm{K}}}$, the $j$th entry of $\bm{\lambda}_{\G}$ is updated as $[\bm{\lambda}^{\star}_{\G}]_{[j]} = \max \big\{ {1}/{\mu} - {1}/{[\bm{\lambda}_{\tilde{\bm{K}}}]_{[j]} }   ,0 \big\}$,
where $\mu$ is the real-valued water-level found using, e.g., the bisection method so that $\sum_{j = 1} [\bm{\lambda}^{\star}_{\G}]_{[j]}= 1$. Finally, we calculate $\G_{\text{ERR-ANOMAX}} = \bm{U}_{\G} \text{diag}(\bm{\lambda}^{\star}_{\G}) \bm{V}^{H}_{\G}$, which is then normalized so that $\|\G_{\text{ERR-ANOMAX}}  \x\|^2_{2} = P_{{\RS}}$.

\vspace{-10pt}
\section{HAD-based design}\label{HADRS}
\vspace{-5pt}

As mentioned in \cite{Molisch2017}, the fully-digital system is impractical in mm-wave massive MIMO systems due to the large number of RF chains. Therefore, we consider a HAD AF structure with $N_{\RS} < M_{\RS}$ RF chains, so that the amplification matrix $\G_{k}$ has a structure given as $\G_{k} \overset{\textrm{def}}{=} \A_{\Tx} \B_{k} \A^{T}_{\Rx}$,
where $ \B_{k}$ is the $N_{\RS} \times N_{\RS}$ baseband beamforming matrix for the $k$th subcarrier, while $\A_{\Tx}$ and $\A_{\Rx} $ are the common $M_{\RS} \times N_{\RS}$ transmit and received analog beamforming matrices, respectively, implemented using phase-shifting networks. Therefore, we have that $|[\A_{\Tx}]_{[i,j]}| = 1, \forall i,j,$ and $|[\A_{\Rx}]_{[i,j]}| = 1, \forall i,j$. 
Considering the HAD amplification matrix given above, the maximization of the total SE of the system in (\ref{eq:SE}) is more challenging due to the constant modulus constraints of the analog beamforming matrices. Therefore, we propose a heuristic solution to (\ref{eq:SE}), similarly to \cite{Ayach2012}, as \begingroup\makeatletter\def\f@size{9.5}\check@mathfonts
\def\maketag@@@#1{\hbox{\m@th\small\normalfont#1}}%
\begin{equation}\label{Prob1}
	\begin{aligned}
		{\min_{ \A_{\Tx}, \A_{\Rx}  , \B_{k}, \forall k}} &\textstyle \sum_{k = 0}^{K - 1} \Big\Vert \bm{G}_{\FD,k} -  \A_{\Tx} \B_{k} \A^{T}_{\Rx} \Big\Vert_{F}^2 \\
		& |[\A_{\Rx}]_{[i,j]}| = 1,  \forall i,j,  |[\A_{\Tx}]_{[i,j]}| = 1, \forall i,j,
	\end{aligned}
\end{equation} \endgroup
where $\bm{G}_{\FD,k}$ is a given unconstrained fully-digital RS amplification matrix on the $k$th subcarrier.
For the known $\bm{G}_{\FD,k},\forall k$, we form a 3-way tensor $\Gt_{\FD}$ by concatenating $\bm{G}_{\FD,k},\forall k$, along the third mode, as
\begingroup\makeatletter\def\f@size{9.5}\check@mathfonts
\def\maketag@@@#1{\hbox{\m@th\small\normalfont#1}}%
\begin{equation}\label{FDTen}
	\Gt_{\FD} = [\G_{\FD,0}  \sqcup_3 \dots \sqcup_3 \G_{\FD,K - 1} ] \in \mathbb{C}^{M_\RS \times M_\RS \times K}. 
\end{equation}
\endgroup
Here, we note that the tensor $\Gt_{\FD}$ in (\ref{FDTen}) admits a Tucker2 decomposition \cite{Tucker1966,HOSVDLathauwer,Cichocki2015} given as 
\begingroup\makeatletter\def\f@size{9.5}\check@mathfonts
\def\maketag@@@#1{\hbox{\m@th\small\normalfont#1}}%
\begin{align}\label{Tens}
	\Gt_{\FD} = \Bt \times_1 \A_{\Tx} \times_2 \A_{\Rx} \times_3 \I_{K},
\end{align}
\endgroup
where $\Bt \in \CC^{N_{\RS} \times N_{\RS} \times K } $ is the core tensor formed as $ \Bt = [\B_{0}\sqcup_3 \dots \sqcup_3 \B_{K-1} ]$. Utilizing (\ref{Tens}), (\ref{Prob1}) can be rewritten as a constrained Tucker2 decomposition as 
\begingroup\makeatletter\def\f@size{9.5}\check@mathfonts
\def\maketag@@@#1{\hbox{\m@th\small\normalfont#1}}%
\begin{equation}
	\begin{aligned}
		{\min_{ \A_{\Tx}, \A_{\Rx} , \Bt }} & \quad \| \Gt_{\FD}  -  \Bt \times_1 \A_{\Tx} \times_2 \A_{\Rx}  \times_3 \I_{K}\|_{\Fro}^2\\
		& |[\A_{\Rx}]_{[i,j]}| = 1,  \forall i,j, \text{ and  }
		|[\A_{\Tx}]_{[i,j]}| = 1, \forall i,j.
	\end{aligned}
	\label{eq:AmpTucker2}
\end{equation} 
\endgroup

\textbf{HOSVD-based solution}: A direct solution to (\ref{eq:AmpTucker2}) can be obtained from the HOSVD \cite{HOSVDLathauwer} of the tensor $\Gt_{\FD}$. Specifically, let the HOSVD of $\Gt_{\FD}$ be written as 
\begingroup\makeatletter\def\f@size{9.5}\check@mathfonts
\def\maketag@@@#1{\hbox{\m@th\small\normalfont#1}}%
\begin{equation}
	\Gt_{\FD} = \bm{\mathcal{S}} \times_1 \U_1 \times_2 \U_2 \times_3 \U_{3},
\end{equation} \endgroup
where $\U_1$, $\U_2$, and $\U_{3} $ are the unitary factor matrices, while $ \bm{\mathcal{S}}$ is the core tensor. Then, by utilizing the fact that  $\A_{\Tx} $ (resp. $ \A_{\Rx}$) spans the same subspace as $\U_1$ (resp. $\U_2$), the analog matrices and the baseband matrix of the $k$th subcarrier can be obtained as ${\mathring{\A}_{\Tx}} = \bm{\Pi}([\U_1]_{[1:N_\RS]} )$, ${\mathring{\A}_{\Rx}} = \bm{\Pi}([\U_2]_{[1:N_\RS]} )$, and $\mathring{\B}_{k} = [\bm{\mathcal{S}}]_{[1:N_\RS,1:N_\RS,k]}$, where $\bm{\Pi}(\cdot)$ is an element-wise projection function defined as $\bm{\Pi}(u) = u / |u|$. Therefore, the $k$th HAD RS amplification matrix is given as $\G_{\textrm{HOSVD},k} = {\mathring{\A}_{\Tx}} {\mathring{\B}_{k}} {\mathring{\A}^{T}_{\Rx}}$, which is then normalized so that $\|\G_{\textrm{HOSVD},k} \x_k \|^2_{2} = P_{{\RS}}$.

\textbf{AltMax-based solution}: Another solution to (\ref{eq:AmpTucker2}) can be obtained using our recently proposed AltMax approach in \cite{Gherekhloo2020}. Specifically, let $[\Gt_{\FD}]_{(1)} = \A_{\Tx} [\Bt]_{(1)} (\bm{I}_{K} \otimes \A_{\Rx} )^{T}$ denote the         one-mode unfolding of $\Gt_{\FD} $,
in which $[\Bt]_{(1)}$ is formed according to the forward cyclical ordering \cite{Naskovska2018}. Then, a solution to $\A_{\Tx}$ can be obtained as
\begingroup\makeatletter\def\f@size{9.5}\check@mathfonts
\def\maketag@@@#1{\hbox{\m@th\small\normalfont#1}}%
\begin{equation}\label{tracemax}
	\begin{aligned}
		\max_{ \A_{\Tx} }  & \quad \Vert \A^{H}_{\Tx} [\Gt_{\FD}]_{(1)}  \Vert^2_{F} = 	\max_{ \A_{\Tx}  } \text{trace}(\A^{H}_{\Tx} \bm{G}_{(1)} \A_{\Tx} ) \\
		\text{s.t.} & \quad |[\A_{\Tx}]_{[i,j]}| = 1, \forall i,j,
	\end{aligned} 
\end{equation} \endgroup
where $\bm{G}_{(1)} = [\Gt_{\FD}]_{(1)} [\Gt_{\FD}]^{H}_{(1)}$. Our proposed AltMax technique in \cite{Gherekhloo2020} can be used to solve (\ref{tracemax}), which updates $\A_{\Tx}$ column-wise until a convergence is reached. Similarly to (\ref{tracemax}), the $\A_{\Rx}$ matrix can be calculated utilizing the two-mode unfolding $[\Gt_{\FD}]_{(2)} = \A_{\Rx} [\Bt]_{(2)} (\bm{I}_{K} \otimes \A_{\Tx} )^{T}$. 

From the obtained $\dot{\A}_{\Tx}$ and $\dot{\A}_{\Rx}$, the core tensor can be calculated as $[\dot{\Bt}]_{(3)} = [\Gt_{\FD}]_{(3)} [(\dot{\A}_{\Rx} \otimes \dot{\A}_{\Tx} )^{T}]^+$,
where $[\Gt_{\FD}]_{(3)} = [\Bt]_{(3)} (\A_{\Rx} \otimes \A_{\Tx} )^{T}$. Thus, the $k$th HAD RS amplification matrix is given as $\G_{\textrm{AltMax},k} = {\dot{\A}_{\Tx}} {\dot{\B}_{k}} {\dot{\A}^{T}_{\Rx}}$, where ${\dot{\B}_{k}}$ is the $k$th 3-mode slice of $\dot{\Bt}$. Finally, we normalize $\G_{\textrm{AltMax},k}$ so that $\|\G_{\textrm{AltMax},k} \x_k \|^2_{2} = P_{{\RS}}$.

\begin{figure}[t!]
	\centering
	\begin{subfigure}[t]{0.5\linewidth}
		\centering
		\includegraphics[width=1\linewidth]{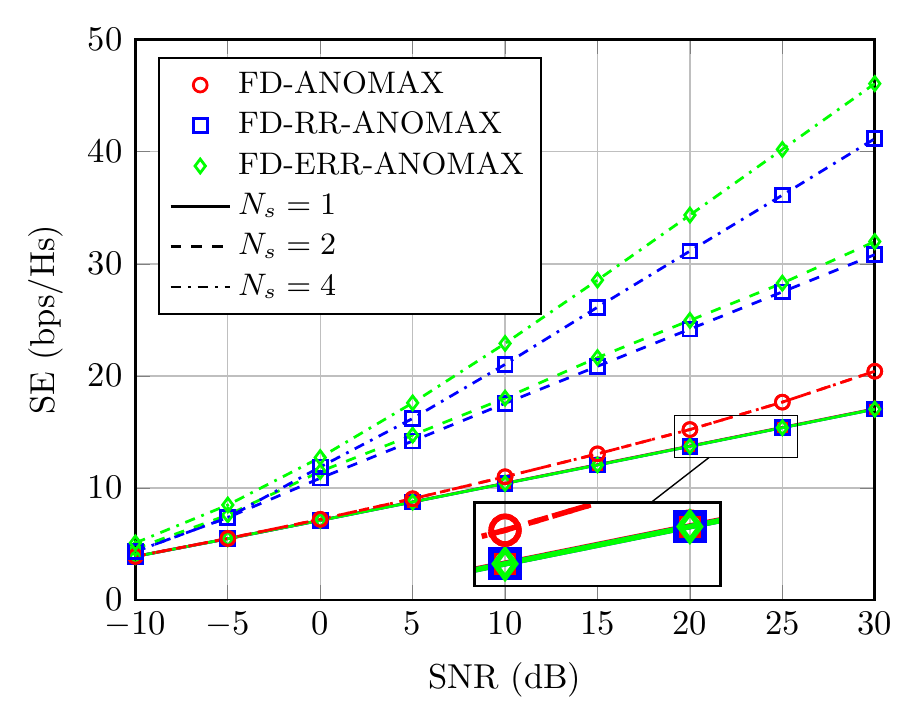}
		\caption{{\scriptsize SE vs. SNR [$R = 2$]}}
	\end{subfigure}%
	~ 
	\begin{subfigure}[t]{0.475\linewidth}
		\centering
		\includegraphics[width=1\linewidth]{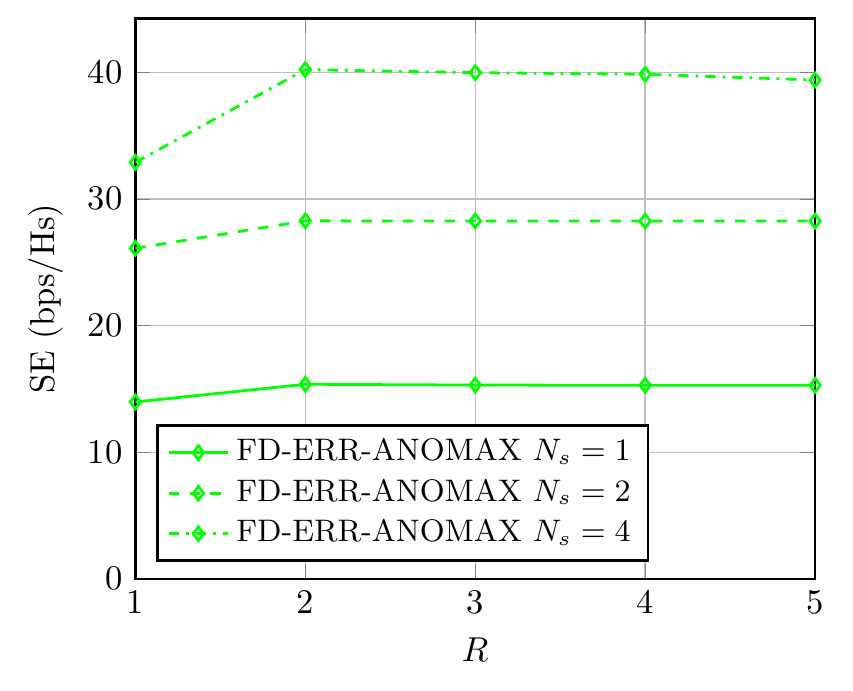}
		\caption{{\scriptsize SE vs. $R$ [SNR = 25dB]}}
	\end{subfigure}
	\caption{Example 1: The fully-digital case [$K = 1$]}
	\label{fig:fig1}
		\vspace{-10pt}
\end{figure}

\begin{figure}[t!]
	\centering
	\begin{subfigure}[t]{0.48\linewidth}
		\centering
		\includegraphics[width=1\linewidth]{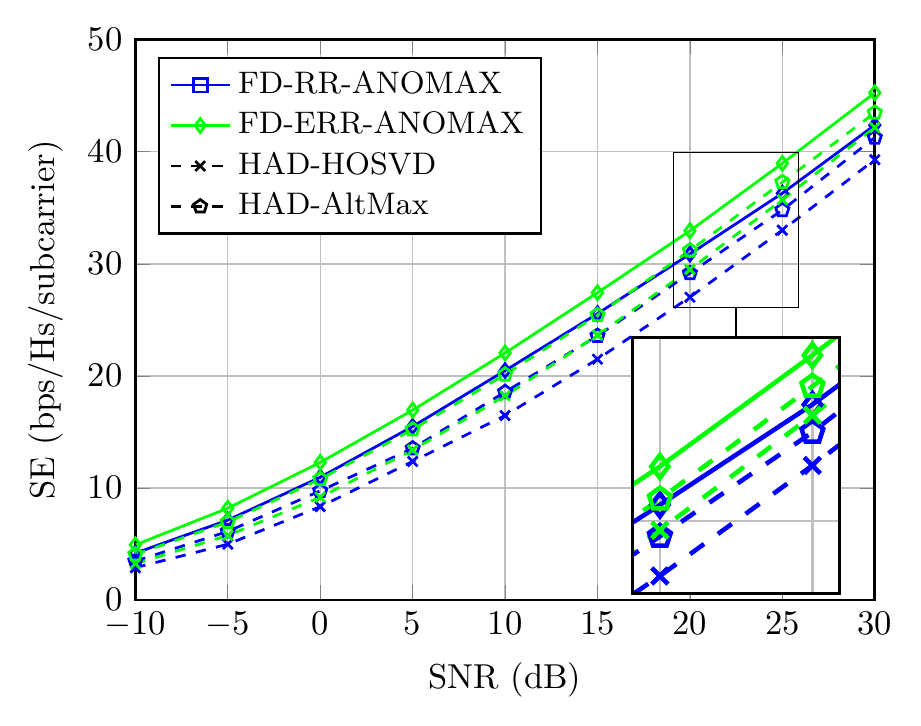}
		\caption{{\scriptsize  SE vs. SNR [$N_\RS = 8,K = 32$]}}
		 		\vspace{-15pt}
	\end{subfigure}%
	~ 
	\begin{subfigure}[t]{0.48\linewidth}
		\centering
		\includegraphics[width=1\linewidth]{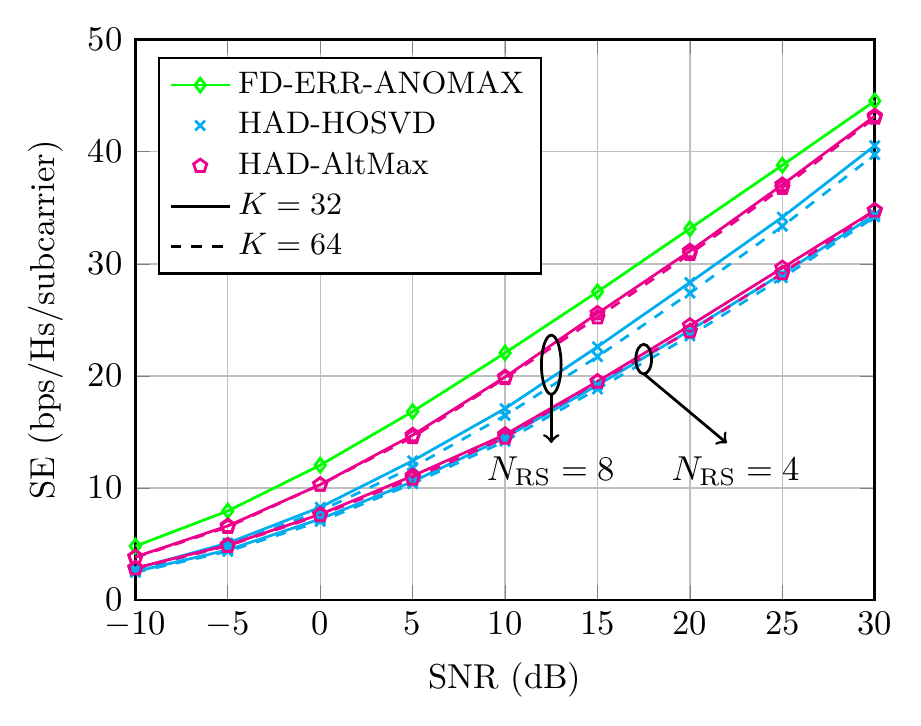}
		\caption{{\scriptsize SE vs. SNR}}
	\end{subfigure}
	\caption{Example 2: The HAD case [$R = 2, N_s = 4$]}
	\label{fig:fig2}
		\vspace{-10pt}
\end{figure}

\vspace{-10pt}
\section{Simulation Results}\label{numresults}
\vspace{-10pt}
Similarly to \cite{Gherekhloo2020}, we assume that each channel matrix is generated using the classical Saleh-Valenzuela model \cite{SV}, where we fix the number of channel paths $L = 6$. We define the signal-to-noise ratio as $\text{SNR} = \frac{1}{\sigma^2}$, where we assume that $P_\RS = P_\UE = 1$ and $\sigma^2_{\RS} = \sigma^2_{\UE} = \sigma^2$. In all the simulation scenarios, we use $M_\RS = 64 $ and $ M_1 = M_2 = 4$.  

\textbf{Example 1: The fully-digital case:} Here, we compare the performance of ANOMAX \cite{Roemer2009}, RR-ANOMAX \cite{Roemer2010b}, and the proposed ERR-ANOMAX assuming that $K = 1$ and the RS is equipped with a fully-digital beamforming structure. 

From Fig. \ref{fig:fig1}(a), we can see that when $N_s = 1$, the three methods achieve the same performance. However, the rank restoration methods, i.e., RR-ANOMAX and ERR-ANOMAX have a significant performance gain in the $N_s > 1$ scenarios when compared to ANOMAX. For $N_s > 1$, ERR-ANOMAX outperforms the RR-ANOMAX of \cite{Roemer2010b}, specially for $N_s = 4$. Here, in Fig. \ref{fig:fig1}(a), we have assumed that $R = 2$ for ERR-ANOMAX. Using computer simulations, as reported in Fig. \ref{fig:fig1}(b), we have observed that setting $ R = 2$ provides the best performance.  

\textbf{Example 2: The HAD case:} Here, we compare RR-ANOMAX and the proposed ERR-ANOMAX assuming that the RS is equipped with a HAD beamforming structure. Fig.~\ref{fig:fig2} shows the SE versus the SNR.

From Fig. \ref{fig:fig2}(a), as expected, we can observe that ERR-ANOMAX maintains its advantages over RR-ANOMAX in the HAD beamforming scenarios as well, where the AltMax method is shown to outperform the HOSVD method, agreeing with our results reported in \cite{Gherekhloo2020}. However, from Fig. \ref{fig:fig2}(b), we can see that when $N_\RS$ reduces from $8$ to $4$, the performance of the HAD methods reduces significantly, where both HAD methods seem to have an equal performance. This implies that, for the HAD architectures, the number of RF chains $N_\RS$ at the RS should be equal or larger than the total number of data streams, i.e., $N_\RS \geq 2N_s$ to maintain a close performance to that achieved by the fully-digital counterpart. Finally, Fig. \ref{fig:fig2}(b) also shows that when the number of system subcarriers $K$ increases from $32$ to $64$, the performance of the HAD methods slightly decreases, which is more evident with the HOSVD method and less with the AltMax method.

\vspace{-10pt}
\section{Conclusions}
\vspace{-10pt}
In this work, we have enhanced and extended the RR-ANOMAX scheme, originally proposed in \cite{Roemer2010b} for single-carrier FD AF MIMO TWR systems, to multi-carrier HAD AF MIMO-OFDM TWR systems. Specifically, we have proposed the ERR-ANOMAX scheme that outperforms RR-ANOMAX in multi-stream communications. Moreover, we have shown that the HAD amplification matrix design can be formulated as a constrained Tucker2 decomposition, for which two solutions are proposed, an HOSVD-based solution and an AltMax-based solution. Simulation results show that ERR-ANOMAX AltMax outperforms the other methods.

	\newpage
	\bibliographystyle{IEEEtran}
	\bibliography{References}
	
\end{document}